\documentclass[prc,aps,twocolumn,showpacs]{revtex4}
\usepackage{amsmath,amssymb}
\usepackage{graphicx}
\def\pcomma{$^,$}
\def\pmainz{$^1$}
\def\pglsg{$^2$}
\def\pkent{$^3$}
\def\pbonn{$^4$}
\def\pgatch{$^5$}
\def\pgiess{$^6$}
\def\ppavia{$^7$}
\def\pgwu{$^8$}
\def\pucla{$^9$}
\def\plpi{$^{10}$}
\def\pdalh{$^{11}$}
\def\psaint{$^{12}$}
\def\pbasel{$^{13}$}
\def\ptomsk{$^{14}$}
\def\pedinb{$^{15}$}
\def\psackv{$^{16}$}
\def\plund{$^{17}$}
\def\pinr{$^{18}$}
\def\pzagreb{$^{19}$}
\def\pcua{$^{20}$}
\begin{document}

\title{
 Measurement of the $\gamma p\to K^0\Sigma^+$ reaction with the Crystal Ball/TAPS detectors
 at the Mainz Microtron}

\author{
P.~Aguar-Bartolom\'e\pmainz,
J.~R.~M.~Annand\pglsg,
H.~J.~Arends\pmainz,
K.~Bantawa\pkent,
R.~Beck\pbonn,
V.~Bekrenev\pgatch,
H.~Bergh\"auser\pgiess,
A.~Braghieri\ppavia,
W.~J.~Briscoe\pgwu,
J.~Brudvik\pucla,
S.~Cherepnya\plpi,
R.~F.~B.~Codling\pglsg,
C.~Collicott\pdalh\pcomma\psaint,
B.~Demissie\pgwu,
M.~Dieterle\pbasel,
E.~J.~Downie\pmainz\pcomma\pgwu,
P.~Drexler\pgiess,
L.~V.~Fil'kov\plpi,
A.~Fix\ptomsk,
D.~I.~Glazier\pedinb,
R.~Gregor\pgiess,
D.~Hamilton\pglsg,
E.~Heid\pmainz\pcomma\pgwu,
D.~Hornidge\psackv,
D.~Howdle\pglsg,
L.~Isaksson\plund,
I.~Jaegle\pbasel,
O.~Jahn\pmainz,
T.~C.~Jude\pedinb,
V.~L.~Kashevarov\pmainz\pcomma\plpi,
I.~Keshelashvili\pbasel,
R.~Kondratiev\pinr,
M.~Korolija\pzagreb,
M.~Kotulla\pgiess,
A.~Koulbardis\pgatch,
S.~Kruglov\pgatch,
B.~Krusche\pbasel,
V.~Lisin\pinr,
K.~Livingston\pglsg,
I.~J.~D.~MacGregor\pglsg,
Y.~Maghrbi\pbasel,
J.~Mancel\pglsg,
D.~M.~Manley\pkent,
E.~F.~McNicoll\pglsg,
D.~Mekterovic\pzagreb,
V.~Metag\pgiess,
A.~Mushkarenkov\ppavia,
B.~M.~K.~Nefkens\pucla,
A.~Nikolaev\pbonn,
R.~Novotny\pgiess,
M.~Oberle\pbasel,
H.~Ortega\pmainz,
M.~Ostrick\pmainz\footnote[1]{corresponding author; e-mail: ostrick@kph.uni-mainz.de},
P.~Ott\pmainz,
P.~B.~Otte\pmainz,
B.~Oussena\pmainz\pcomma\pgwu,
P.~Pedroni\ppavia,
F.~Pheron\pbasel,
A.~Polonski\pinr,
S.~Prakhov\pgwu\pcomma\pucla\footnote[2]{corresponding author; e-mail: prakhov@ucla.edu},
J.~Robinson\pglsg,
G.~Rosner\pglsg,
T.~Rostomyan\pbasel,
S.~Schumann\pmainz,
M.~H.~Sikora\pedinb,
D.~I.~Sober\pcua,
A.~Starostin\pucla,
I.~I.~Strakovsky\pgwu,
I.~M.~Suarez\pucla,
I.~Supek\pzagreb,
C.~M.~Tarbert\pedinb,
M.~Thiel\pgiess,
A.~Thomas\pmainz,
M.~Unverzagt\pmainz\pcomma\pbonn,
D.~P.~Watts\pedinb,
D.~Werthm\"uller\pbasel,
L.~Witthauer\pbasel,
and F.~Zehr\pbasel
 \\
\vspace*{0.1in}
(A2 Collaboration at MAMI)
\vspace*{0.1in}
}

\affiliation{
\pmainz Institut f\"ur Kernphysik, Johannes Gutenberg-Universit\"at Mainz,
D-55099 Mainz, Germany}
\affiliation{
\pglsg SUPA, School of Physics and Astronomy, University of Glasgow, Glasgow G12 8QQ, 
United Kingdom}
\affiliation{
\pkent Kent State University, Kent, Ohio 44242-0001, USA}
\affiliation{
\pbonn Helmholtz-Institut f\"ur Strahlen- und Kernphysik, University of Bonn, 
D-53115 Bonn, Germany}
\affiliation{
\pgatch Petersburg Nuclear Physics Institute, 188350 Gatchina, Russia}
\affiliation{
\pgiess II Physikalisches Institut, University of Giessen, D-35392 Giessen, Germany}
\affiliation{
\ppavia INFN Sesione di Pavia, I-27100 Pavia, Italy}
\affiliation{
\pgwu The George Washington University, Washington, DC 20052-0001, USA}
\affiliation{
\pucla University of California Los Angeles, Los Angeles, California 90095-1547, USA}
\affiliation{
\plpi Lebedev Physical Institute, 119991 Moscow, Russia}
\affiliation{
\pdalh Dalhousie University, Halifax, Nova Scotia B3H 4R2, Canada}
\affiliation{
\psaint Saint Mary's University, Halifax, Nova Scotia B3H 3C3, Canada}
\affiliation{
\pbasel Institut f\"ur Physik, University of Basel, CH-4056 Basel, Switzerland}
\affiliation{
\ptomsk Laboratory of Mathematical Physics, Tomsk Polytechnic University, 634050 Tomsk, Russia}
\affiliation{
\pedinb SUPA, School of Physics, University of Edinburgh, Edinburgh EH9 3JZ, United Kingdom}
\affiliation{
\psackv Mount Allison University, Sackville, New Brunswick E4L 1E6, Canada}
\affiliation{
\plund Lund University, SE-22100 Lund, Sweden}
\affiliation{
\pinr Institute for Nuclear Research, 125047 Moscow, Russia}
\affiliation{
\pzagreb Rudjer Boskovic Institute, HR-10000 Zagreb, Croatia}
\affiliation{
\pcua The Catholic University of America, Washington, DC 20064, USA}

\date{\today}
                  
\begin{abstract}
 The $\gamma p\to K^0\Sigma^+$ reaction has been measured from threshold to $E_{\gamma}=1.45$~GeV
 ($W_{\mathrm{CM}}=1.9$~GeV) using the Crystal Ball and TAPS multiphoton
 spectrometers together with the photon tagging facility at the Mainz Microtron MAMI.
 In the present experiment, this reaction was searched for in the $3\pi^0 p$
 final state, by assuming $K^0_S \to \pi^0\pi^0$ and $\Sigma^+\to \pi^0 p$.
 The experimental results include total and differential cross sections
 as well as the polarization of the recoil hyperon. 
 The new data significantly improve empirical knowledge about the 
 $\gamma p\to K^0\Sigma^+$ reaction in the measured energy range.
 The results are compared to previous measurements and model predictions. 
 It is demonstrated that adding the present $\gamma p\to K^0\Sigma^+$ results
 to existing data allowed a better description of this reaction
 with various models.
 
\end{abstract}


\maketitle

\section{Introduction}

The unique extraction of partial-wave scattering amplitudes
and universal baryon-resonance parameters 
from experimental data as well as their precise interpretation
in QCD ranks among the most challenging tasks 
in modern hadron physics. 
During the last decades, an enormous effort  
to study baryon resonances in photoinduced meson production at 
various laboratories has started. 
Many single- and double-spin observables for different final states 
have been measured for the first time. 

Photoinduced kaon-hyperon ($KY$) production has attracted much 
attention for two reasons.
First, open strangeness production necessarily involves the generation 
of a strange $q \bar{q}$ pair, and new aspects of nucleon spectroscopy might be
discovered, which are not manifest in reactions containing pion-nucleon 
initial or final states. Second, the weak hyperon decays 
reveal the polarization state of the final-state baryon. This information is 
essential in order to achieve a model-independent amplitude reconstruction in
a so-called complete experiment~\cite{barker, sandorfi}. 

The basic reactions include $\gamma p \to K^+ \Lambda$, $\gamma p \to K^+ \Sigma^0$,
and $\gamma p \to K^0 \Sigma^+$, as well as the corresponding reactions off the neutron. 
The production of an isoscalar $\Lambda$ hyperon involves only $N^*$ resonances, while
$\gamma p\to K^+\Sigma^0$ and $\gamma p\to K^0\Sigma^+$ 
may receive contributions from both $N^*$ and $\Delta^*$ states.
Therefore, an experimentally based isospin
decomposition of the transition amplitudes requires data on both
reactions with similar quality.

Differential cross sections and hyperon polarization for reactions with a 
charged kaon, $\gamma p \to K^+ \Lambda$ and $\gamma p \to K^+ \Sigma^0$,
were measured at Jefferson Lab using the CLAS
 detector~\cite{Kpl_CLAS_2004,Kpl_CLAS_2006,Kpl_CLAS_2010},
 at ELSA using the SAPHIR detector~\cite{Kpl_SAPHIR_1998,Kpl_SAPHIR},
 at GRAAL~\cite{Kpl_GRAAL_2007}, and with LEPS~\cite{Kpl_LEPS_2006,Kpl_LEPS2_2006}. 
 Recent measurements of the beam-recoil observables with CLAS~\cite{br-clas}
 and GRAAL~\cite{br-graal}
 are important steps toward a model-independent amplitude reconstruction. 

The $\gamma p\to K^0\Sigma^+$ reaction can be identified via the
decays $K^0_S \to \pi^0\pi^0$ or $K^0_S \to \pi^+\pi^-$
and $\Sigma^+ \to \pi^0 p$ or $\Sigma^+ \to \pi^+ n$.
Differential cross sections have been measured by CBELSA/TAPS,
using the $\pi^0\pi^0\pi^0 p$ final state~\cite{K0Spl_CBELSA_2008,K0Spl_CBELSA_2010},
and by SAPHIR~\cite{K0Spl_SAPHIR_1999,K0Spl_SAPHIR} and CLAS~\cite{K0Spl_CLAS},
using the $\pi^+\pi^-\pi^0 p$ and $\pi^+\pi^-\pi^+ n$ final states.  
Compared to the $K^+ \Lambda$ and $K^+ \Sigma^0$ channels, the statistical accuracy
of the existing $K^0\Sigma^+$ data are rather limited, especially at energies below 
$E_{\gamma}=1.5$~GeV. 

In parallel with these experimental efforts, a lot of theoretical studies for 
modeling and understanding the dynamics of kaon photoproduction were started. 
Calculations have been performed in the framework of single- or 
multichannel isobar models~\cite{K0Spl_KMAID, RPR_2007, BGPWA},
coupled-channel approaches~\cite{giessen, scholten, juelich} 
as well as chiral-unitary approaches~\cite{Chiral}.
It turned out that, in contrast to single $\pi$ or $\eta$ photoproduction,
both resonant and nonresonant contributions 
play an almost equally important part in the reaction dynamics. 
Therefore, extracted resonance parameters strongly depend on the treatment
of the nonresonant contributions and a conclusive picture about the dynamics and 
the baryon-resonance contributions has not yet been achieved.

In the phenomenological KAON-MAID model~\cite{K0Spl_KMAID},
the nonresonant part of the amplitudes contains $s-$, $u-$, and $t-$channel 
terms along with a contact interaction, which is required to restore gauge invariance 
after hadronic form factors are introduced. The model
was fitted to the first SAPHIR $\gamma p\to K^+\Lambda$,
 $\gamma p\to K^+\Sigma^0$~\cite{Kpl_SAPHIR_1998},
 and $\gamma p\to K^0\Sigma^+$ data~\cite{K0Spl_SAPHIR_1999}. 
In the Regge-Plus-Resonance (RPR) approach of Refs.~\cite{RPR_2007, RPR_A}, 
modeling nonresonant contributions via exchange of Regge trajectories
involves only a few parameters, which can be constrained by
the high-energy data above the resonance region.
The latest Bonn-Gatchina (BnGa) partial-wave analysis 
(PWA)~\cite{BGPWA} incorporates the majority of existing data from pion- 
 and photoinduced reactions into simultaneous multichannel fits.
In Ref.~\cite{Chiral}, a unitary, gauge-invariant coupled-channel
approach based on a chiral effective Lagrangian was used to describe the threshold
regions of kaon photoproduction without including any resonances at all.

 This work reports on new experimental results for the
 $\gamma p\to K^0\Sigma^+$ reaction that include total and differential cross sections
 and recoil polarization of $\Sigma^+$. The data were obtained with the 
 Crystal Ball/TAPS multiphoton spectrometers, using the
 energy-tagged beam of bremsstrahlung photons from the Mainz Microtron.
 Compared to the previous experiments, the present statistics and the quality of the data
 allow us to measure the $\gamma p\to K^0\Sigma^+$ reaction at $E_{\gamma}<1450$~MeV
 with a higher accuracy and smaller energy binning.

\section{Experimental setup}
\label{sec:Setup}

The reaction $\gamma p\to K^0\Sigma^+$
was measured using the Crystal Ball (CB)~\cite{CB}
as the central spectrometer and TAPS~\cite{TAPS,TAPS2}
as a forward spectrometer. These detectors were
installed in the energy-tagged bremsstrahlung photon beam of
the Mainz Microtron (MAMI)~\cite{MAMI,MAMIC}. 
The photon energies are determined
by the Glasgow tagging spectrometer~\cite{TAGGER2,TAGGER,TAGGER1}.

The CB detector is a sphere consisting of 672
optically isolated NaI(Tl) crystals, shaped as
truncated triangular pyramids, which point toward
the center of the sphere. The crystals are arranged in two
hemispheres that cover 93\% of $4\pi$ sr, sitting
outside a central spherical cavity with a radius of
25~cm, which is designed to hold the target and inner
detectors. In this experiment, TAPS was
arranged in a plane consisting of 384 BaF$_2$
counters of hexagonal cross section. It was
installed 1.5~m downstream of the CB center
and covered the full azimuthal range for polar angles
from $1^\circ$ to $20^\circ$.
More details on the energy and angular resolution of the CB and TAPS
are given in Refs.~\cite{slopemamic,etamamic}.
\begin{figure*}
\includegraphics[width=15.5cm,height=5.7cm,bbllx=1.cm,bblly=.0cm,bburx=19.cm,bbury=5.2cm]{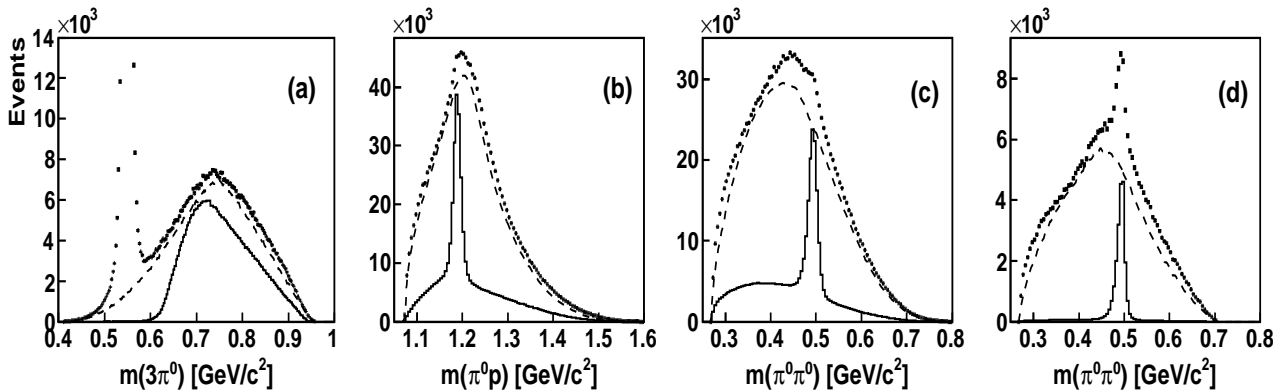}
\caption{
 Comparison of the invariant-mass distributions (a)~$m(3\pi^0)$,
 (b)~$m(\pi^0p)$ , and (c) and (d)~$m(\pi^0\pi^0)$ , obtained
 for the experimental $\gamma p\to 3\pi^0 p$ events (crosses) and
 for the MC simulation of $\gamma p\to K^0\Sigma^+ \to 3\pi^0 p$ (solid line)
 and $\gamma p\to 2\pi^0 \Delta^+(1232)\to 3\pi^0 p$ (dashed line).
 All distribution are obtained for events above the $\gamma p\to K^0\Sigma^+$ threshold.
 The $m(\pi^0p)$ and $m(\pi^0\pi^0)$ distributions are obtained for
 $m(3\pi^0)>0.6$~GeV/$c^2$. The events in (d) are selected by testing
 the $\gamma p\to \pi^0\pi^0\Sigma^+\to 3\pi^0 p$ hypothesis.
}
 \label{fig:m3pi0p} 
\end{figure*}

The present measurement used 1508- and 1557-MeV
electron beams from the upgraded Mainz Microtron, MAMI-C~\cite{MAMIC}.
The data with the 1508-MeV beam were taken in 2007 and those with the 1557-MeV beam
in 2009. Bremsstrahlung photons, produced by the 1508-MeV electrons
in a 10-$\mu$m Cu radiator and collimated by a 4-mm-diameter Pb collimator,
 were incident on a 5-cm-long liquid hydrogen (LH$_2$) target located
in the center of the CB. The energies of the incident
photons were analyzed up to 1402~MeV by detecting
the post-bremsstrahlung electrons in the Glasgow tagger~\cite{TAGGER2}.
With the 1557-MeV electron beam, the incident photons were analyzed up to 1448~MeV,
and a 10-cm-long LH$_2$ target was used.
The energy resolution of the tagged photons is mostly defined by the width
of tagger focal-plane detectors and by the electron-beam energy.
For the present beam energies, a typical width of a tagger channel was about 4~MeV.
Due to the beam collimation, only part of the bremsstrahlung photon flux
reached the LH$_2$ target. In order to extract the reaction cross sections,
the probability of bremsstrahlung photons reaching the target 
(the so-called tagging efficiency) was measured for each tagger channel.
The average tagging efficiency in the experiment with the 1508-MeV electron beam
was found to vary from 69\% at the $\gamma p\to K^0\Sigma^+$ threshold to 64\%
at $E_{\gamma}=1.4$~GeV.
With the 1557-MeV electron beam, the average tagging efficiency was similar
to the 1508-MeV-beam magnitudes up to $E_{\gamma}=1.4$~GeV, dropping then to 57\%
at $E_{\gamma}=1.45$~GeV.

The experimental trigger in the measurement with the 1508-MeV electron beam
required two conditions: 1. the total-energy deposited in the CB had
to exceed $\sim$320~MeV and 2. the number of so-called hardware clusters
in the CB (multiplicity trigger) had to be larger than two.
With the 1557-MeV electron beam, the trigger on the total energy
in the CB was increased to $\sim$340~MeV.
TAPS was not in the multiplicity trigger for these experiments.

More details on the experimental conditions of the data taking with the 1508-MeV electron beam
in 2007 are given in Refs.~\cite{slopemamic,etamamic}.

\section{Data handling}\label{Data}

 The reaction $\gamma p\to K^0\Sigma^+$ was searched for in the $3\pi^0 p$
 final state, by assuming $K^0_S \to \pi^0\pi^0$ and $\Sigma^+\to \pi^0 p$.
 The $\gamma p\to 3\pi^0 p \to 6\gamma p$ candidates were extracted from
 the analysis of events having
 six and seven clusters reconstructed in the CB and TAPS together.
 As reported in Ref.~\cite{etamamic}, a large part of the $3\pi^0 p$ final
 state comes from the process $\gamma p \to \eta p \to 3\pi^0 p$.
 However, this process can be easily separated from $\gamma p\to K^0_S\Sigma^+$
 based on the invariant mass of the $3\pi^0$ system,
 which is heavier than the $\eta$ mass
 for $\gamma p\to K^0_S\Sigma^+$ events.
 Another part of $\gamma p\to 3\pi^0 p$ is the so-called direct $3\pi^0$
 production that comes from processes
 involving only a chain of $N^*$ and $\Delta^*$ decays. 
\begin{figure*}
\includegraphics[width=15.cm,height=6.cm,bbllx=1.cm,bblly=.5cm,bburx=19.cm,bbury=8.cm]{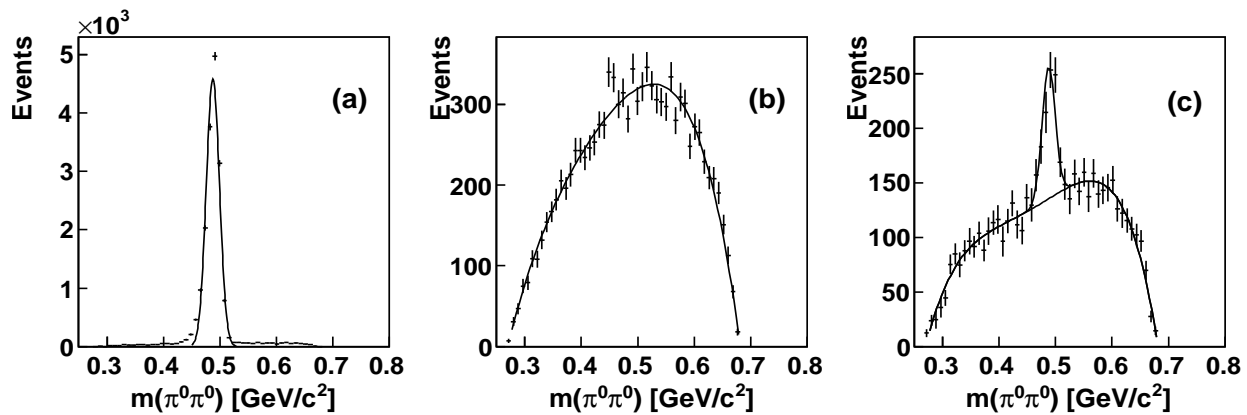}
\caption{
 The $m(\pi^0\pi^0)$ distributions for
 ~(a) the MC simulation of $\gamma p\to K^0\Sigma^+\to 3\pi^0 p$ with a Gaussian fit,
 ~(b) the MC simulation of $\gamma p\to 2\pi^0 \Delta^+(1232)\to 3\pi^0 p$
    with a fit of a polynomial of order five, and
 ~(c) the experimental data fitted with the sum of a Gaussian and a polynomial of order five. 
}
 \label{fig:m2pi0_fit} 
\end{figure*}

 Besides the physical $3\pi^0$ background, there are two more background sources.
 The first one comes from interactions of
 incident photons in the windows of the target cell.
 The subtraction of this background from
 the experimental spectra was based on the 
 analysis of the data samples that were taken
 with an empty (no liquid hydrogen) target.
 Another background is due to random coincidences
 of the tagger hits with the experimental trigger;
 its subtraction was done by using 
 only those tagger hits for which all coincidences were random
 (see Refs.~\cite{slopemamic,etamamic} for more details).

 The selection of event candidates was based on
 the kinematic-fit technique.
 The details of the kinematic-fit parametrization
 of the detector information and resolution are given
 in Ref.~\cite{slopemamic}.
 The hypothesis $\gamma p \to 3\pi^0 p \to 6\gamma p$ was tested to identify
 all events that have the $3\pi^0 p$ final state.
 The events that satisfied this hypothesis at the 2\% confidence level (CL)
 (i.e., with a probability of misinterpretation less than 2\%)
 were accepted as the reaction candidates.
 The kinematic-fit output for which the pairing combination
 of the six photons to the three $\pi^0$s had the largest CL
 was used to reconstruct the reaction kinematics. 
 The hypothesis $\gamma p \to \eta p \to 3\pi^0 p \to 6\gamma p$ was tested
 to identify those events that were produced via the $\eta \to 3\pi^0$ decay.
 The test of the hypothesis $\gamma p \to K^0\Sigma^+ \to 3\pi^0 p \to 6\gamma p$
 was not done as it would accept all direct $3\pi^0 p$ events that have a combination
 of the invariant masses of the $\pi^0 p$ and $\pi^0\pi^0$ systems
 simultaneously close to the mass of $\Sigma^+$ and $K^0$.

 The determination of the experimental acceptance was based on
 a Monte Carlo (MC) simulation of the $\gamma p \to K^0\Sigma^+ \to 3\pi^0 p$
 reaction with an isotropic production angular distribution.
 Two background reactions were simulated as $\gamma p \to \eta p \to 3\pi^0 p$
 and $\gamma p\to 2\pi^0 \Delta^+(1232)\to 3\pi^0 p$,
 also with isotropic production angular distributions.
 For the data at each electron-beam energy, the corresponding MC events
 were propagated through a {\sc GEANT} (version 3.21) simulation of the experimental
 setup. To reproduce resolutions of the experimental data,
 the {\sc GEANT} output (energy and timing) was subject
 to additional smearing, thus allowing both the simulated and experimental data
 to be analyzed in the same way.
 The simulated events were also tested for whether they passed the trigger requirements.
\begin{figure*}
\includegraphics[width=15.cm,height=10.cm,bbllx=.5cm,bblly=.5cm,bburx=19.cm,bbury=11.5cm]{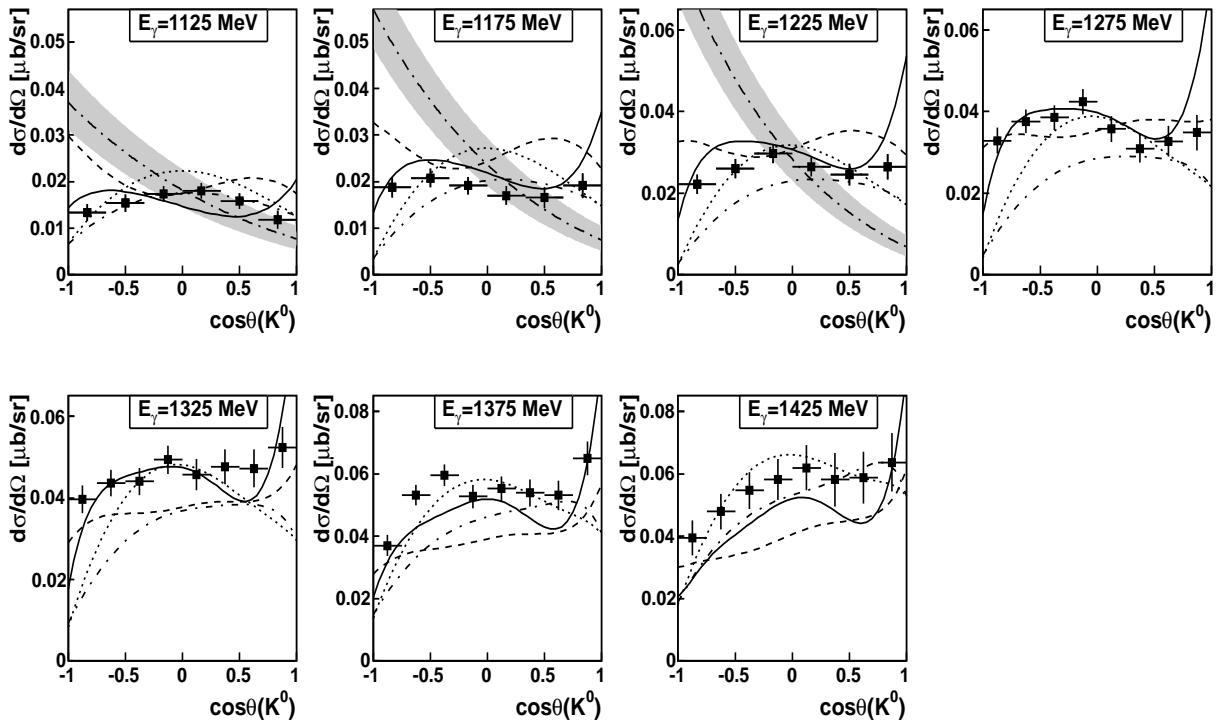}
\caption{
 $\gamma p\to K^0\Sigma^+$ differential cross sections of this work  (solid squares) compared to
 the predictions of various models: BG2011-2 (solid line), BG2011-1 (dashed line)~\protect\cite{BGPWA},
 RPR-3 (dotted line)~\protect\cite{RPR_2007}, RPR-A (dash-dotted line)~\protect\cite{RPR_A},
 and chiral-unitary approach (long-dash-dotted line with gray error band)~\protect\cite{Chiral}. 
}
 \label{fig:dxs_k0sigp_theor} 
\end{figure*}

 In Fig.~\ref{fig:m3pi0p}, various invariant-mass distributions obtained
 from the experimental $\gamma p\to 3\pi^0 p$ events are compared to those 
 obtained from the MC simulation of the processes
 $\gamma p\to K^0\Sigma^+ \to 3\pi^0 p$ and $\gamma p\to 2\pi^0 \Delta^+(1232)\to 3\pi^0 p$.
 All events are plotted only for beam energies above the $\gamma p\to K^0\Sigma^+$ threshold
 ($\sim$1048~MeV).
 The $3\pi^0$ invariant-mass distributions are depicted in Fig.~\ref{fig:m3pi0p}(a).
 As seen, the experimental distribution includes a strong peak from
 $\gamma p \to \eta p \to 3\pi^0 p$ events and a structure that is similar to
 the MC simulation of $\gamma p\to 2\pi^0 \Delta^+(1232)\to 3\pi^0 p$.
 Also, one can see that events from $\gamma p\to K^0\Sigma^+\to 3\pi^0 p$ and
 $\gamma p \to \eta p \to 3\pi^0 p$ barely overlap.
 The $\pi^0 p$ invariant-mass distributions are shown in Fig.~\ref{fig:m3pi0p}(b)
 for events with $m(3\pi^0)>0.6$~GeV/$c^2$.
 The corresponding invariant mass for the $\pi^0\pi^0$ system is shown
 in Fig.~\ref{fig:m3pi0p}(c).
 These distributions for $m(\pi^0 p)$ and $m(\pi^0\pi^0)$ have three entries
 per one event because of the number of final-state $\pi^0$s.
 As seen in Fig.~\ref{fig:m3pi0p}(b), the experimental $m(\pi^0 p)$ spectrum
 reveals a clear signal from the $\Delta^+(1232)\to \pi^0 p$ decay,
 the shape of which is close to the MC simulation
 of $\gamma p\to 2\pi^0 \Delta^+(1232)\to 3\pi^0 p$.  
 As a weak signal from $\Sigma^+(1189)\to \pi^0 p$ sits on a big bump from
 $\Delta^+(1232)\to \pi^0 p$, this makes it difficult to use the $m(\pi^0 p)$
 distribution for fitting the $\Sigma^+$ signal above background.
 As seen in Fig.~\ref{fig:m3pi0p}(c), the $\pi^0\pi^0$ background
 under the $K^0_S\to \pi^0\pi^0$ signal is much smoother. The shape
 of this background is also close to the MC simulation of
 $\gamma p\to 2\pi^0 \Delta^+(1232)\to 3\pi^0 p$.
 To enlarge the signal-to-background ratio in the $m(\pi^0\pi^0)$ distribution,
 only events from the region of the $\Sigma^+(1189)$ peak
 should be accepted into the $m(\pi^0\pi^0)$ distribution.
 The width of this peak can be determined from the MC simulation of
 $\gamma p\to K^0\Sigma^+\to 3\pi^0 p$, which is shown in Fig.~\ref{fig:m3pi0p}(b)
 by a solid line. Such an approach was used in
 the previous analyses of the CBELSA data~\cite{K0Spl_CBELSA_2008,K0Spl_CBELSA_2010}.
 The present analysis revealed that an even better solution is to test
 the hypothesis $\gamma p\to \pi^0\pi^0\Sigma^+\to 3\pi^0 p$ with
 the kinematic fit. Besides selecting events
 from the region of the $\Sigma^+(1189)$ peak, the use of this hypothesis
 also improves the angular resolution and the $m(\pi^0\pi^0)$ resolution
 for the actual $\gamma p\to K^0\Sigma^+\to 3\pi^0 p$
 events. Fitting the $\gamma p\to \pi^0\pi^0\Sigma^+\to 3\pi^0 p$ hypothesis
 resulted in $\sigma=3.4^\circ$ for the resolution in the $K^0_S$ polar angle
 in the center-of-mass (c.m.) frame and $\sigma=8.6$~MeV/$c^2$ for
 the invariant-mass resolution of the two $\pi^0$ from the $K^0_S$ decay,
 whereas the corresponding values based on the fit of
 the $\gamma p\to 3\pi^0 p$ hypothesis were $3.6^\circ$ and $11.0$~MeV/$c^2$.
 The results of fitting the $\gamma p\to \pi^0\pi^0\Sigma^+\to 3\pi^0 p$
 hypothesis are shown in Fig.~\ref{fig:m3pi0p}(d).
 The $m(\pi^0\pi^0)$ distribution from the MC simulation of
 $\gamma p\to K^0\Sigma^+\to 3\pi^0 p$ includes now an almost clean, narrower peak
 from $K^0_S\to \pi^0\pi^0$. The $K^0_S$ signal in the experimental $m(\pi^0\pi^0)$
 distribution is much more prominent than the one in Fig.~\ref{fig:m3pi0p}(c).

 All previous measurements of $\gamma p\to K^0\Sigma^+$ were made with
 wide energy bins; the smallest bins were 100-MeV wide.
 This prevented an adequate characterization of the region from threshold
 to the excitation-function maximum, causing difficulties with the theoretical
 interpretation of the data.
 The present data were divided into eight energy bins
 from $E_\gamma=1050$~MeV to $E_\gamma=1450$~MeV, allowing 50-MeV energy binning
 for measuring both the $\gamma p\to K^0\Sigma^+$ cross sections
 and recoil polarization of $\Sigma^+$.  
 The data within one energy bin were divided
 into six $\cos\theta$ bins for $E_\gamma<1250$~MeV and into eight $\cos\theta$ bins
 for $E_\gamma>1250$~MeV, where $\theta$ is the angle between the direction of 
 the outgoing $K^0$ and the incident photon in the c.m. frame.   

 Although the 50-MeV binning covers entirely the energy range available in this experiment,
 unfortunately, the energy points themselves do not match any existing data.
 For a better comparison of the present and previous data,
 an additional set of results was obtained also with the use of 50-MeV bins,
 but with the energy points matching all previous measurements.
 Namely, the additional set includes five energy points:
 1100, 1200, 1250, 1300, and 1400~MeV.
 This additional set was used only for illustration
 (Figs.~\ref{fig:dxs_k0sigp_exp} and~\ref{fig:rpl_k0sigp_exp})
 and cannot be used in any further analysis,
 as its results are strongly correlated with the results of the main set. 

 The $K^0_S\to \pi^0\pi^0$ signal was fitted individually
 in every $\cos\theta$ bin. The fitting procedure is illustrated
 in Fig.~\ref{fig:m2pi0_fit} for one particular energy and
 $\cos\theta$ bin.
 The $m(\pi^0\pi^0)$ distribution from
 the MC simulation of $\gamma p\to K^0\Sigma^+\to 3\pi^0 p$ fitted with a Gaussian
 is shown in Fig.~\ref{fig:m2pi0_fit}(a). The $m(\pi^0\pi^0)$ distribution
 from the MC simulation of $\gamma p\to 2\pi^0 \Delta^+(1232)\to 3\pi^0 p$
 fitted with a polynomial of order five is shown in Fig.~\ref{fig:m2pi0_fit}(b).
 The experimental $m(\pi^0\pi^0)$ distribution fitted with the sum of a Gaussian
 and a polynomial of order five is shown in Fig.~\ref{fig:m2pi0_fit}(c).
 For the latter fit, the initial parameters for the polynomial were taken from the
 previous fit to the MC simulation of $\gamma p\to 2\pi^0 \Delta^+(1232)\to 3\pi^0 p$,
 and two parameters of the Gaussian (namely, the mean value and $\sigma$)
 were fixed to the values from the previous fit to the MC simulation
 of $\gamma p\to K^0\Sigma^+\to 3\pi^0 p$.  
 Since the calculation of the experimental number of $\gamma p\to K^0\Sigma^+\to 3\pi^0 p$
 events was based on the area under the Gaussian,  
 the corresponding detection efficiency was calculated
 in the same way (based on the area under the Gaussian fit to the MC simulation
 of $\gamma p\to K^0\Sigma^+\to 3\pi^0 p$).
\begin{table*}
\caption
[tab:dxs6bn]{
 Differential cross sections for $\gamma p\to K^0\Sigma^+$ at the energies for which
 the data were divided into six $\cos\theta$ bins.
 } \label{tab:dxs6bn}
\begin{ruledtabular}
\begin{tabular}{|c|c|c|c|} 
 $E_{\gamma}$~[MeV] & $1125\pm 25$ & $1175\pm 25$ & $1225\pm 25$ \\
\hline
 $\cos\theta(K^0)$ & $d\sigma/d\Omega$ [$\mu$b/sr] & $d\sigma/d\Omega$ [$\mu$b/sr] & $d\sigma/d\Omega$ [$\mu$b/sr] \\
\hline
 -1.0 to -0.667  & $0.0133\pm 0.0018$ & $0.0188\pm 0.0022$ & $0.0221\pm 0.0023$ \\
 -0.667 to -0.333  & $0.0154\pm 0.0019$ & $0.0208\pm 0.0020$ & $0.0260\pm 0.0023$ \\
 -0.333 to 0.0 & $0.0174\pm 0.0017$ & $0.0191\pm 0.0019$ & $0.0296\pm 0.0023$ \\
 0.0 to 0.333  & $0.0180\pm 0.0017$ & $0.0170\pm 0.0020$ & $0.0265\pm 0.0023$ \\
 0.333 to 0.667  & $0.0158\pm 0.0017$ & $0.0166\pm 0.0023$ & $0.0246\pm 0.0027$ \\
 0.667 to 1.0  & $0.0118\pm 0.0020$ & $0.0191\pm 0.0026$ & $0.0265\pm 0.0031$ \\
\end{tabular}
\end{ruledtabular}
\end{table*}
\begin{table*}
\caption
[tab:dxs8bn]{
 Differential cross sections for $\gamma p\to K^0\Sigma^+$ at the energies for which
 the data were divided into eight $\cos\theta$ bins.
 } \label{tab:dxs8bn}
\begin{ruledtabular}
\begin{tabular}{|c|c|c|c|c|} 
 $E_{\gamma}$~[MeV] & $1275\pm 25$ & $1325\pm 25$ & $1375\pm 25$ & $1425\pm 25$ \\
\hline
  $\cos\theta(K^0)$ & $d\sigma/d\Omega$ [$\mu$b/sr] &
  $d\sigma/d\Omega$ [$\mu$b/sr]&  $d\sigma/d\Omega$ [$\mu$b/sr] &  $d\sigma/d\Omega$ [$\mu$b/sr] \\
\hline
 -1.0 to -0.75  & $0.0329\pm 0.0032$ & $0.0396\pm 0.0034$ & $0.0371\pm 0.0034$ & $0.0395\pm 0.0056$ \\
 -0.75 to -0.5  & $0.0376\pm 0.0029$ & $0.0437\pm 0.0033$ & $0.0532\pm 0.0033$ & $0.0480\pm 0.0057$ \\
 -0.5 to -0.25  & $0.0385\pm 0.0030$ & $0.0441\pm 0.0033$ & $0.0596\pm 0.0035$ & $0.0547\pm 0.0060$ \\
 -0.25 to 0.0  & $0.0424\pm 0.0031$ & $0.0494\pm 0.0035$ & $0.0528\pm 0.0038$ & $0.0582\pm 0.0066$ \\
 0.0 to 0.25  & $0.0357\pm 0.0031$ & $0.0457\pm 0.0038$ & $0.0553\pm 0.0039$ & $0.0620\pm 0.0073$ \\
 0.25 to 0.5  & $0.0309\pm 0.0035$ & $0.0476\pm 0.0042$ & $0.0540\pm 0.0044$ & $0.0581\pm 0.0086$ \\
 0.5 to 0.75  & $0.0327\pm 0.0038$ & $0.0472\pm 0.0046$ & $0.0531\pm 0.0047$ & $0.0587\pm 0.0083$ \\
 0.75 to 1.0  & $0.0348\pm 0.0043$ & $0.0524\pm 0.0051$ & $0.0650\pm 0.0054$ & $0.0637\pm 0.0094$ \\
\end{tabular}
\end{ruledtabular}
\end{table*}

 The recoil polarization $P$ of the $\Sigma^+$ hyperon
 was measured via its parity-violating weak decay $\Sigma^+\to \pi^0 p$.
 The angular distribution of the decay nucleon is given by Ref.~\cite{Gatto} as
\begin{equation}
 W(\cos\xi_N) \sim 1+\alpha P \cos\xi_N~,
\label{eqn:wang}
\end{equation}
 where $\alpha=-0.98$ is the asymmetry parameter for the $\Sigma^+\to \pi^0 p$ decay.
 The angle $\xi$ is defined by 
\begin{equation}
\label{eqn:xiang}
 \cos\xi = (\hat{\boldsymbol{\gamma}} \times \hat{\boldsymbol{K}}^0)
 \cdot \hat{\boldsymbol{n}} /
 |\hat{\boldsymbol{\gamma}} \times \hat{\boldsymbol{K}}^0|
 = \hat{\boldsymbol{N}} \cdot \hat{\boldsymbol{n}}~,
\end{equation}
 where $\hat{\boldsymbol{\gamma}}$ and $\hat{\boldsymbol{K}}^0$
 are, respectively, vectors in the direction of the incident photon and
 the outgoing $K^0$, $\hat{\boldsymbol{n}}$ is a unit vector in the direction of
 the decay proton in the $\Sigma^+$ rest frame,
 and $\hat{\boldsymbol{N}}$ is the normal to the production plane.
 Based on the decay asymmetry of $\Sigma^+$ with respect to the production plane,
 the polarization was measured by averaging the proton angular distribution
 above ($\cos\xi>0$) and below ($\cos\xi<0$) this plane:
\begin{equation}
 P = \frac{2}{\alpha} \frac{N_{\mathrm{up}}-N_{\mathrm{down}}}{N_{\mathrm{up}}+N_{\mathrm{down}}}~,
\label{eqn:pol}
\end{equation}
 where $N_{\mathrm{up}}$ and $N_{\mathrm{down}}$ represent, respectively, the number of events
 with the decay proton emitted above and below the production plane.

\section{Discussion of the results}

 The differential cross sections for $\gamma p\to K^0\Sigma^+$ and 
 recoil polarization of $\Sigma^+$ were obtained
 as a function of $\cos\theta_{K^0}$, where $\theta_{K^0}$ is the angle
 between the direction of $K^0$ and the incident photon in the c.m. frame.
 As discussed in Sec.~\ref{Data},
 the main results were obtained for eight 50-MeV intervals,
 covering the full energy range available in the experiment.
 Additional results were obtained just for a better comparison
 with existing data, also using 50-MeV intervals,
 but with the energy points matching the previous measurements.

 Since the mean value and $\sigma$ of the Gaussian were fixed while
 fitting to the experimental $m(\pi^0\pi^0)$ spectra,
 the statistical uncertainties in the results for the differential cross sections
 and recoil polarization are based on the fit errors for the Gaussian height
 parameter. These uncertainties are shown in the figures and listed in the tables.

 The systematic uncertainties in the results are more essential for the
 differential cross sections, and most of them are canceled in the
 calculation of recoil polarization. 
 The largest contribution to the systematic
 uncertainties comes from the parametrization of the background in the fits.
 According to our study, this uncertainty is about 0.003~$\mu$b/sr
 for each individual value in the differential cross sections,
 independent of its magnitude. This uncertainty
 was estimated by changing the degree of a polynomial used in the fit and
 the $m(\pi^0\pi^0)$ range fitted. Although for the largest values
 in the differential cross sections, this uncertainty is only about 5\%,
 for the smallest values, it reaches 25\%.
 The systematic uncertainty because of the combinatorial background
 under the peak from the $K^0_S\to \pi^0\pi^0$ decays (see Fig.~\ref{fig:m2pi0_fit}(a))
 is at the level of 1\%--2\%, depending on the $\cos\theta_{K^0}$ value. 
 The overall systematic uncertainty that comes from the determination of
 the experimental acceptance with the MC simulation and from
 the calculation of the photon-beam flux is about 4\%.
 This uncertainty for the $3\pi^0 p$ final state was studied by
 measuring the process $\gamma p \to \eta p \to 3\pi^0 p$,
 which has a much larger yield and little background,
 for the same data sets (see Ref.~\cite{etamamic} for more details).

 The results below $E_\gamma=1400$~MeV are obtained as a weighted average of the results
 from the data with the 1508-MeV beam and with the 1557-MeV beam.
 The results obtained for the first energy bin, $1050-1100$,
 are not very reliable because of a very weak $K^0\Sigma^+$ signal above large background,
 so they are omitted in this work.
\begin{figure*}
\includegraphics[width=13.cm,height=10.cm,bbllx=1.cm,bblly=.5cm,bburx=19.cm,bbury=16.5cm]{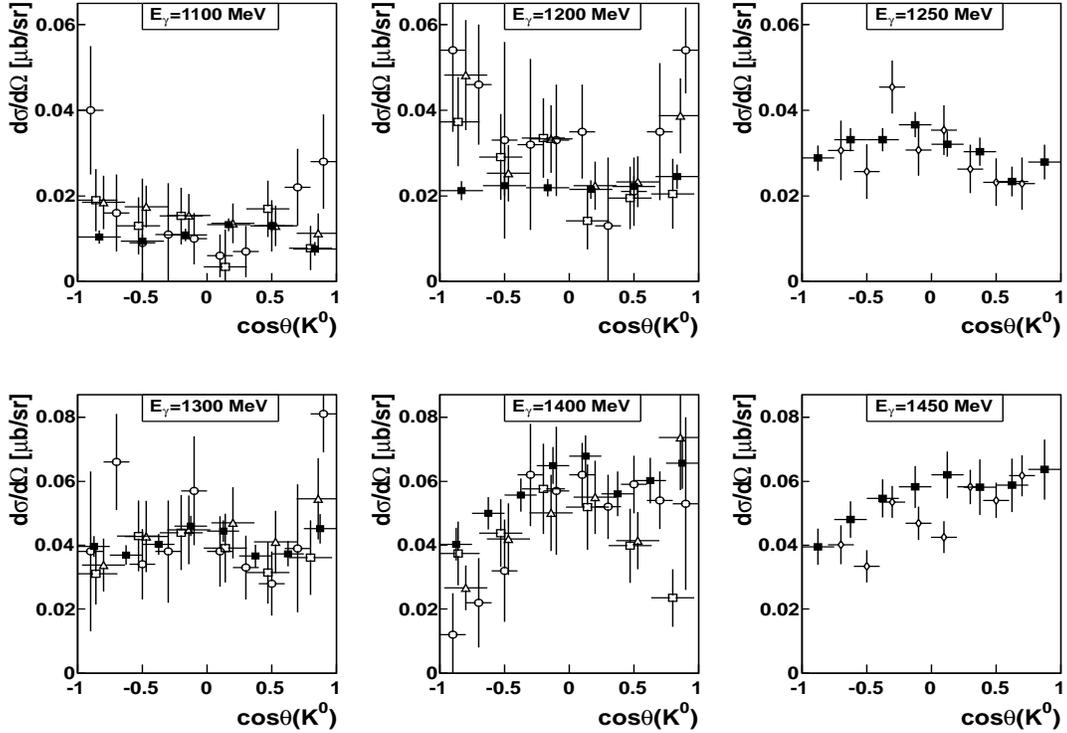}
\caption{
 $\gamma p\to K^0\Sigma^+$ differential cross sections from the additional
 set of this work (solid squares), compared
 to the previous measurements: SAPHIR (open circles, 100-MeV bins)~\protect\cite{K0Spl_SAPHIR},
 CB-ELSA 2008 (open squares, 100-MeV bins)~\protect\cite{K0Spl_CBELSA_2008},
 CB-ELSA 2010 (open triangles, 100-MeV bins)~\protect\cite{K0Spl_CBELSA_2010},
 and CLAS (open diamonds, 200-MeV bins, with CLAS data at 1450~MeV being
 compared to the results of this work at an energy of 1425~MeV)~\protect\cite{K0Spl_CLAS}.  
}
 \label{fig:dxs_k0sigp_exp} 
\end{figure*}
\begin{figure*}
\includegraphics[width=15.cm,height=6.5cm,bbllx=1.cm,bblly=.5cm,bburx=19.cm,bbury=7.cm]{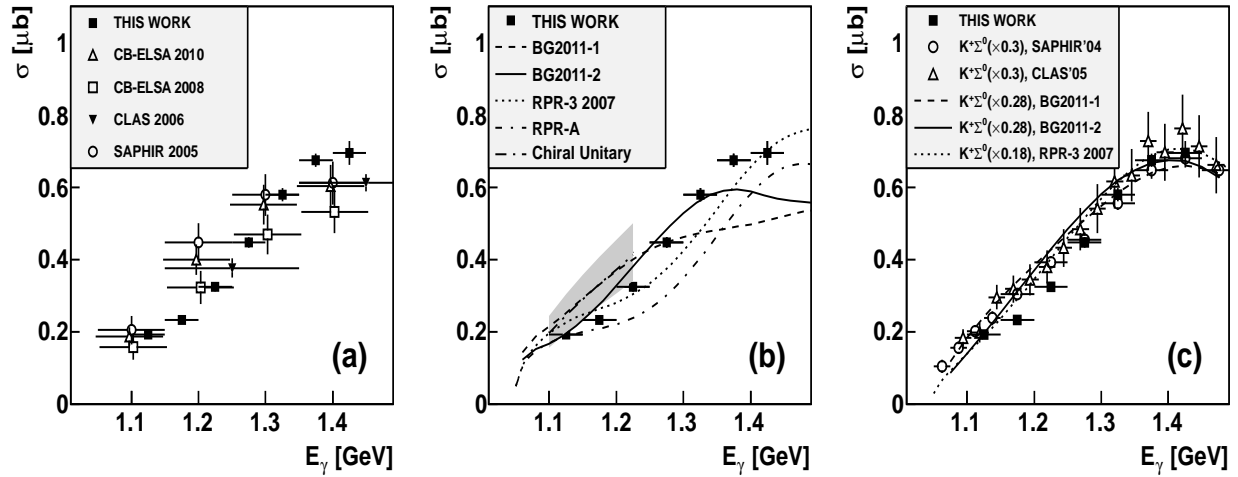}
\caption{
 $\gamma p\to K^0\Sigma^+$ total cross sections of this work (solid squares) compared to
 ~(a) previous measurements by SAPHIR (open circles)~\protect\cite{K0Spl_SAPHIR},
 CLAS (solid triangles)~\protect\cite{K0Spl_CLAS} normalized to the full
 $\cos\theta_{K^0}$ range,
 CB-ELSA 2008 (open squares)~\protect\cite{K0Spl_CBELSA_2008},
 and CB-ELSA 2010 (open triangles)~\protect\cite{K0Spl_CBELSA_2010};
 ~(b) predictions by BG2011-2 (solid line), BG2011-1 (dashed line)~\protect\cite{BGPWA},
 RPR-3 (dotted line)~\protect\cite{RPR_2007}, RPR-A (dash-dotted line)~\protect\cite{RPR_A},
 and the chiral-unitary approach (long-dash-dotted line with gray error band)~\protect\cite{Chiral};
 ~(c) the $\gamma p\to K^+\Sigma^0$ total cross sections by SAPHIR 2004 
 (open circles; multiplied by a factor of 0.3)~\protect\cite{Kpl_SAPHIR},
 CLAS 2006 (open triangles, $\times 0.3$)~\protect\cite{Kpl_CLAS_2006},
 BG2011-2 (solid line, $\times 0.28$), BG2011-1 (dashed line, $\times 0.28$),
 and RPR-3 (dotted line, $\times 0.18$)  
}
 \label{fig:txs_k0sigp_3x1} 
\end{figure*}

 The results of our main set for the $\gamma p\to K^0\Sigma^+$ differential cross sections
 are listed in Tables~\ref{tab:dxs6bn} and~\ref{tab:dxs8bn}.
\begin{figure*}
\includegraphics[width=15.cm,height=10.cm,bbllx=.5cm,bblly=.5cm,bburx=19.cm,bbury=11.5cm]{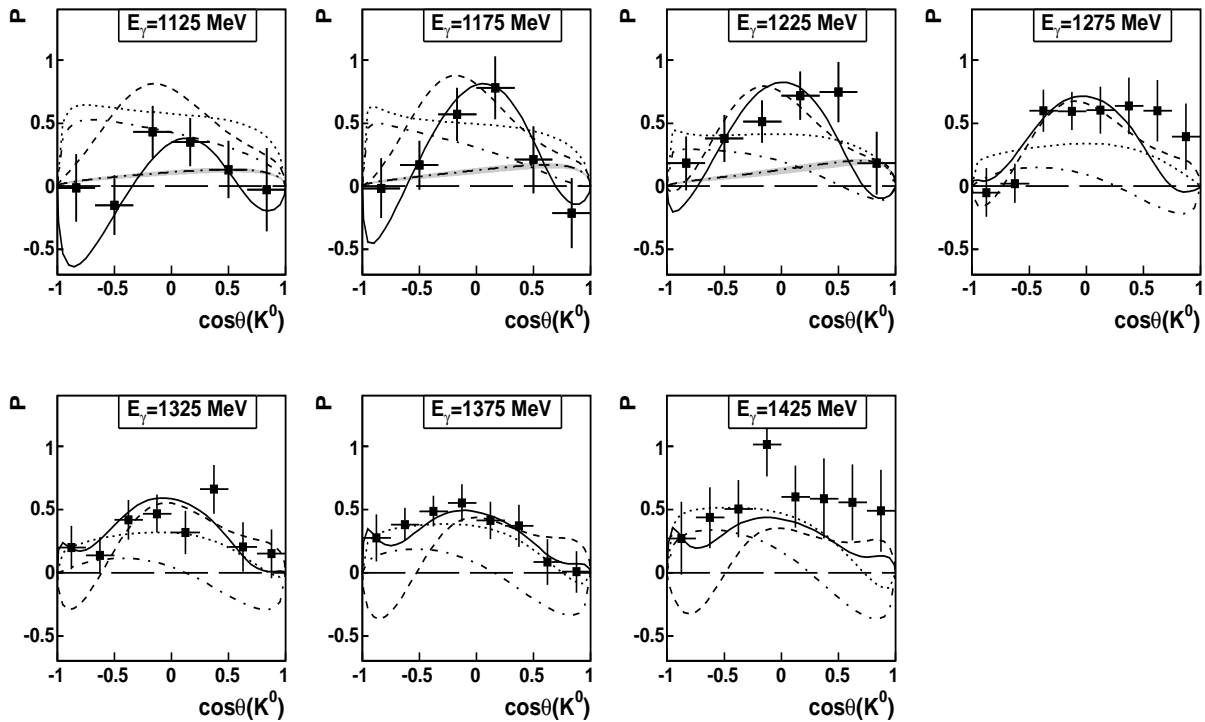}
\caption{
 Results of this work (solid squares) for recoil polarization of $\Sigma^+$ compared to predictions
 from the models BG2011-2 (solid line), BG2011-1 (dashed line), RPR-3 (dotted line), and
 RPR-A (dash-dotted line) and from the chiral-unitary approach (long-dash-dotted line with gray error band). 
}
 \label{fig:rpl_k0sigp_theor} 
\end{figure*}
 In Fig.~\ref{fig:dxs_k0sigp_theor}, 
 these differential cross sections are compared to predictions of various models.
 The latest BnGa multichannel PWA~\cite{BGPWA} permits
 two classes of solutions, called BG2011-1 (shown in the figures by a dashed line) and
 BG2011-2 (shown by a solid line). The RPR approach
 is presented by a solution with background model 3 (RPR-3) from Ref.~\cite{RPR_2007}
 (shown by a dotted line) and a revised solution with background model 3 (RPR-A)
 from Ref.~\cite{RPR_A} (shown by a dash-dotted line).
 The gauge-invariant chiral-unitary approach is shown by a long-dash-dotted line with a 
 gray error band.  None of these predictions is in good agreement with the measured
 differential cross sections for the full energy range.
 The KAON-MAID predictions from Ref.~\cite{K0Spl_KMAID} are not shown here as 
 the model was fitted to the first SAPHIR data~\cite{Kpl_SAPHIR_1998} (not shown here),
 in which the background was significantly underestimated. 
 The chiral-unitary calculation of Ref.~\cite{Chiral} does not describe the angular distributions. 
 However, the order of magnitude of the cross section comes out correctly. This is already
 a success as no baryon resonance contributions are explicitly added in these calculations.
 As seen from the comparison, adding the present results to the existing data will
 require a revision of the analysis of photoinduced $K \Sigma$ production by all models.

 In Fig.~\ref{fig:dxs_k0sigp_exp}, the measured differential cross sections
 from the additional set of this work are compared to all recent data
 in this energy range. As seen,
 all previous measurements are made for wider energy bins and have larger error bars.
 Although the error bars of the other data at $E_{\gamma}=1100$~MeV and $E_{\gamma}=1200$~MeV
 are much larger than the ones of this work,
 the present results appear to be smaller than the previous measurements, especially
 at the extreme angles. This hints that the background in those measurements was
 underestimated at these angles.   
 The results of the CLAS Collaboration~\cite{K0Spl_CLAS}
 at $E_{\gamma}=1250$~MeV and $E_{\gamma}=1450$~MeV
 are in good agreement within the error bars with the results of this work.
 The results obtained at $E_{\gamma}=1300$~MeV and $E_{\gamma}=1400$~MeV are in agreement within
 the error bars with the majority of the previous measurements. 
\begin{table*}
\caption
[tab:rpl6bn]{
 Recoil polarization of $\Sigma^+$ at the energies for which
 the data were divided into six $\cos\theta$ bins.
 } \label{tab:rpl6bn}
\begin{ruledtabular}
\begin{tabular}{|c|c|c|c|} 
 $E_{\gamma}$~[MeV] & $1125\pm 25$ & $1175\pm 25$ & $1225\pm 25$ \\
\hline
 $\cos\theta(K^0)$ & $P$ & $P$ & $P$ \\
\hline
 -1.0 to -0.667  &  $-0.013\pm 0.266$ & $-0.017\pm 0.236$ & $0.181\pm 0.212$ \\
 -0.667 to -0.333  &  $-0.149\pm 0.236$ & $0.167\pm 0.194$ & $0.380\pm 0.185$ \\
 -0.333 to 0.0  &  $0.431\pm 0.207$ & $0.569\pm 0.211$ & $0.513\pm 0.167$ \\
 0.0 to 0.333  & $0.349\pm 0.190$ & $0.780\pm 0.247$ & $0.718\pm 0.191$ \\
 0.333 to 0.667  & $0.133\pm 0.225$ & $0.210\pm 0.265$ & $0.745\pm 0.238$ \\
 0.667 to 1.0  & $-0.026\pm 0.329$ & $-0.212\pm 0.277$ & $0.184\pm 0.250$ \\
\end{tabular}
\end{ruledtabular}
\end{table*}
\begin{table*}
\caption
[tab:rpl8bn]{
 Recoil polarization of $\Sigma^+$ at the energies for which
 the data were divided into eight $\cos\theta$ bins.
 } \label{tab:rpl8bn}
\begin{ruledtabular}
\begin{tabular}{|c|c|c|c|c|} 
 $E_{\gamma}$~[MeV] & $1275\pm 25$ & $1325\pm 25$ & $1375\pm 25$ & $1425\pm 25$ \\
\hline
  $\cos\theta(K^0)$ & $P$ & $P$ & $P$ & $P$ \\
\hline
 -1.0 to -0.75  & $-0.050\pm 0.194$ & $0.199\pm 0.172$ & $0.274\pm 0.186$ & $0.273\pm 0.289$ \\
 -0.75 to -0.5  & $0.023\pm 0.155$ & $0.135\pm 0.147$ & $0.383\pm 0.129$ & $0.436\pm 0.243$ \\
 -0.5 to -0.25  & $0.596\pm 0.167$ & $0.419\pm 0.157$ & $0.486\pm 0.122$ & $0.506\pm 0.226$ \\
 -0.25 to 0.0  & $0.594\pm 0.150$ & $0.469\pm 0.148$ & $0.553\pm 0.149$ & $1.018\pm 0.256$ \\
 0.0 to 0.25  & $0.604\pm 0.186$ & $0.318\pm 0.173$ & $0.414\pm 0.147$ & $0.601\pm 0.246$ \\
 0.25 to 0.5  & $0.637\pm 0.226$ & $0.660\pm 0.192$ & $0.373\pm 0.167$ & $0.588\pm 0.316$ \\
 0.5 to 0.75  & $0.599\pm 0.244$ & $0.204\pm 0.195$ & $0.086\pm 0.181$ & $0.557\pm 0.300$ \\
 0.75 to 1.0  & $0.393\pm 0.264$ & $0.150\pm 0.193$ & $0.007\pm 0.164$ & $0.490\pm 0.325$ \\
\end{tabular}
\end{ruledtabular}
\end{table*}

 The total cross sections for $\gamma p\to K^0\Sigma^+$ were obtained by the integration
 of the present differential cross sections.
 In Fig.~\ref{fig:txs_k0sigp_3x1}(a), these total cross sections
 are compared to the previous measurements. 
 Since the differential cross sections of CLAS were obtained for a limited angular
 range only, their partial total cross sections were normalized by us
 to the ratio of the full and covered ranges of $\cos\theta_{K^0}$.
 As seen, although the previous results have larger error bars, which overlap
 with the present data points, the energy dependence
 of the measured cross sections shows a more steady rise from threshold to the maximum. 
 In Fig.~\ref{fig:txs_k0sigp_3x1}(b), the present total cross sections are compared to
 predictions from the models BG2011-1, BG2011-2, RPR-3, and RPR-A and from the chiral-unitary
 approach. As seen, the prediction from BG2011-2
 is closer to the present data than BG2011-1, and the prediction from RPR-3 is closer than RPR-A.
 However, none of the predictions is sufficiently close to the present data.
 Compared to the previous measurements, the energy dependence
 of the present total cross sections is much closer to the corresponding
 energy dependence of the isospin-related reaction
 $\gamma p\to K^+\Sigma^0$. If both the reactions in this energy range were dominated
 by production via $\Delta^*$ states, the $\gamma p\to K^+\Sigma^0$ yield should
 be about four times as large, compared to $\gamma p\to K^0\Sigma^+$.
 To illustrate this similarity, in Fig.~\ref{fig:txs_k0sigp_3x1}(c)
 the present $\gamma p\to K^0\Sigma^+$ cross sections are compared to the rescaled ones
 of $\gamma p\to K^+\Sigma^0$, which are represented by the CLAS~\cite{Kpl_CLAS_2006}
 and SAPHIR~\cite{Kpl_SAPHIR} data
 and predictions from the models BG2011-1, BG2011-2, and RPR-3.
 As seen, the measured energy dependence for $\gamma p\to K^0\Sigma^+$ is very similar
 to that for $\gamma p\to K^+\Sigma^0$.
\begin{figure}
\includegraphics[width=7.5cm,height=8.5cm,bbllx=1.3cm,bblly=.5cm,bburx=19.cm,bbury=18.cm]{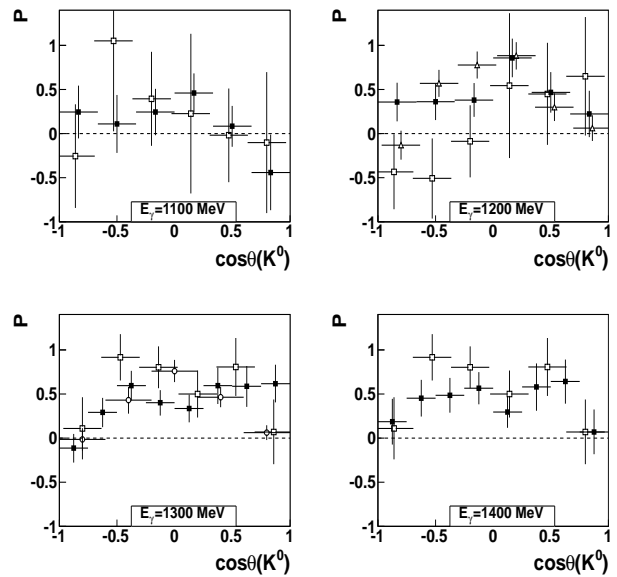}
\caption{
 $\Sigma^+$ recoil-polarization
 results from the additional set of this work (solid squares) compared to the previous
 measurements: SAPHIR (open circles, 250-MeV bins), CB-ELSA 2008 (open squares, 100-MeV bins),
 and CB-ELSA 2010 (open triangles, 150-MeV bins). 
}
 \label{fig:rpl_k0sigp_exp} 
\end{figure}
\begin{figure*}
\includegraphics[width=15.cm,height=10.cm,bbllx=.5cm,bblly=.5cm,bburx=19.cm,bbury=11.5cm]{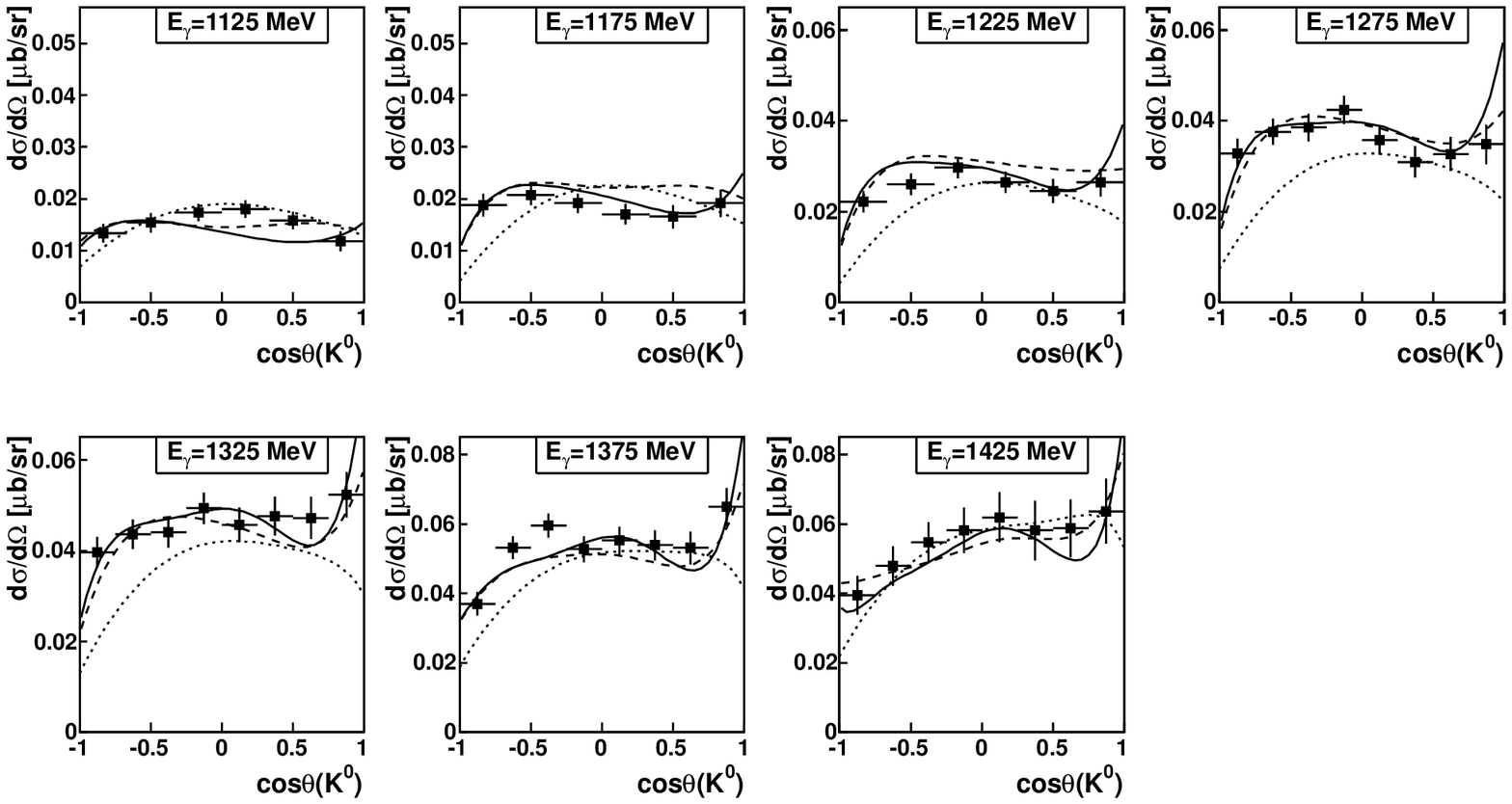}
\caption{
 $\gamma p\to K^0\Sigma^+$ differential cross sections from this work (solid squares)
 compared to the predictions based on new fits of models: BG2011-2 (solid line), BG2011-1 (dashed line),
 and RPR-2007 (dotted line). 
}
 \label{fig:dxs_k0sigp_theor_new} 
\end{figure*}
\begin{figure}
\includegraphics[width=6.5cm,height=7.5cm,bbllx=1.3cm,bblly=.5cm,bburx=17.cm,bbury=18.cm]{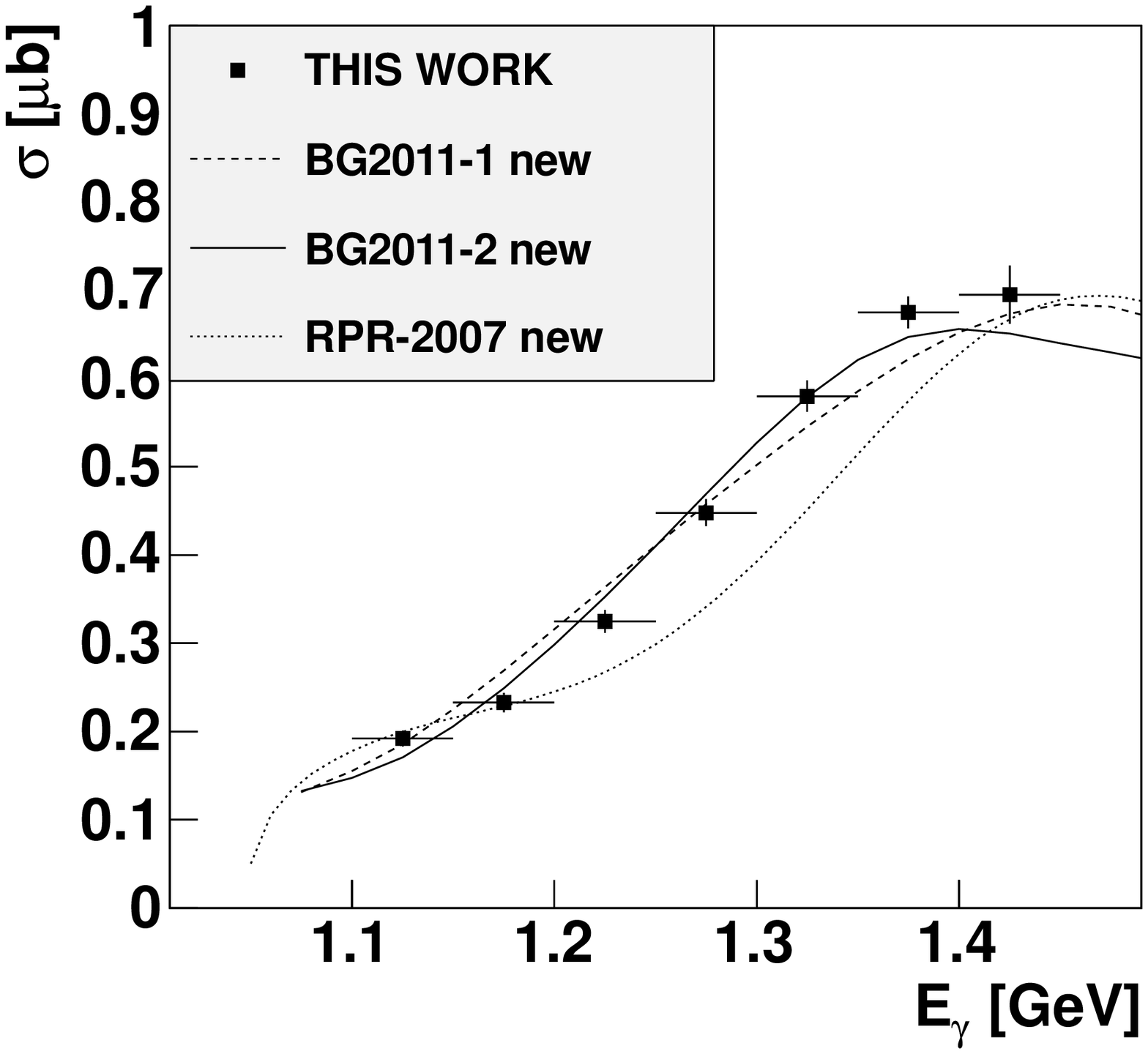}
\caption{
 $\gamma p\to K^0\Sigma^+$  total cross sections from this work (solid squares) compared to
 the predictions based on new fits of models: BG2011-2 (solid line), BG2011-1 (dashed line),
 and RPR-2007 (dotted line).
}
 \label{fig:txs_k0sigp_new} 
\end{figure}
\begin{figure*}
\includegraphics[width=15.cm,height=10.cm,bbllx=.5cm,bblly=.5cm,bburx=19.cm,bbury=11.5cm]{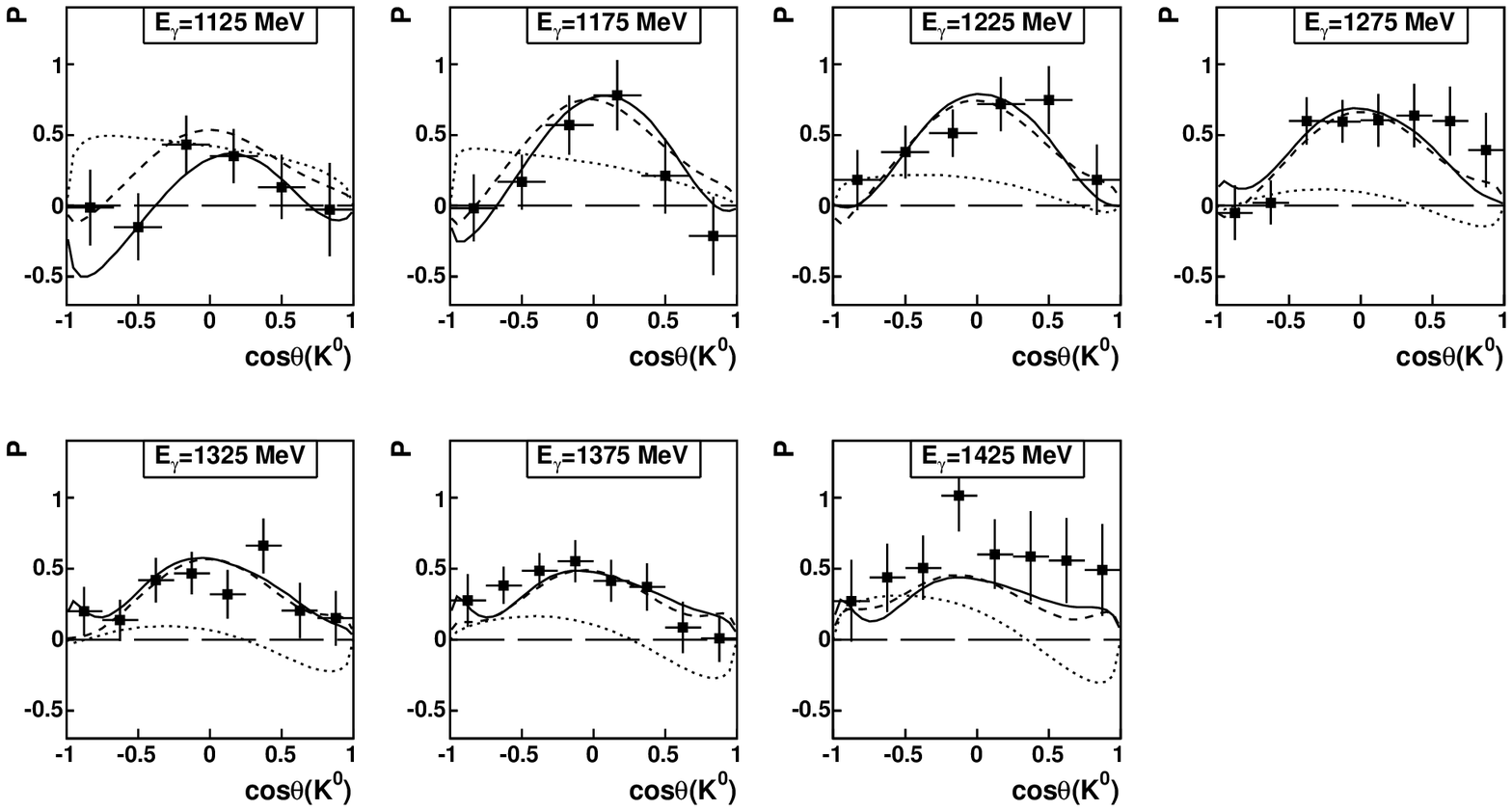}
\caption{
 Results of this work (solid squares) for recoil polarization of $\Sigma^+$ compared
 to predictions based on new fits of models: BG2011-2 (solid line), BG2011-1 (dashed line), and
 RPR-2007 (dotted line).
}
 \label{fig:rpl_k0sigp_theor_new} 
\end{figure*}

 The results of our main set for recoil polarization of $\Sigma^+$
 are listed in Tables~\ref{tab:rpl6bn} and~\ref{tab:rpl8bn}.
 In Fig.~\ref{fig:rpl_k0sigp_theor},  these results
 are compared to predictions from the models BG2011-1, BG2011-2, RPR-3, and RPR-A
 and from the chiral-unitary approach.
 In our opinion, the prediction from BG2011-2 is the closest one for
 most of the data, and the prediction from RPR-3 is better than RPR-A.
 The prediction from the chiral-unitary approach is much smaller than the experimental values.
 In Fig.~\ref{fig:rpl_k0sigp_exp}, the recoil-polarization results from the additional set
 of this work are compared to all previous measurements in this energy range,
 which are represented by only four energies, with wider energy bins
 and larger uncertainties.
 As seen, for all four energies, the present results are in agreement within the error bars
 with the majority of the previous measurements. 

 To check the importance of the present data for a better understanding of
 the $\gamma p\to K^0\Sigma^+$ dynamics in our energy range,
 the RPR and BnGa groups added the results of this work in their analysis
 of $K \Sigma$ photoproduction. The revised results of their fits~\cite{private}
 are shown in Fig.~\ref{fig:dxs_k0sigp_theor_new} for the differential cross sections,
 in Fig.~\ref{fig:txs_k0sigp_new} for the total cross sections, and
 in Fig.~\ref{fig:rpl_k0sigp_theor_new} for recoil polarization of $\Sigma^+$.
 As seen, the RPR-2007 approach, which is a revised RPR-3 model,
 gives a better prediction for the shape of
 the differential cross sections, but leaves a discrepancy for the total cross sections
 and recoil polarization. The RPR-A model was eliminated by the present data. 
 For both the BG2011-1 and BG2011-2 solutions, the updated BnGA PWA 
 yields results which are very close to the present $\gamma p\to K^0\Sigma^+$ data,
 especially for higher photon energies.
 Both RPR and BnGa groups observed a strong influence from the present data on their analyses.
 Details for these results will be published by the RPR and BnGa groups separately~\cite{private}.

\section{Summary and conclusions}\label{conclusion}

 The $\gamma p\to K^0\Sigma^+$ reaction has been measured from threshold to
 $E_{\gamma}=1.45$~GeV. The measurement was conducted
 by using the Crystal Ball and TAPS multiphoton spectrometers
 together with the Glasgow photon tagging facility at the Mainz Microtron.
 The experimental results include total and differential cross sections
 and recoil polarization of $\Sigma^+$.
 The comparison of the present results to previous measurements reveals generally
 good agreement. However, the data of this work are obtained for narrower energy bins
 and with smaller uncertainties. The accuracy of the results presented in this work
 allows a significant improvement of the world data base
 for the $\gamma p\to K^0\Sigma^+$ reaction.
 The comparison of the present results to the existing model predictions indicates
 that none of them describes well the present data in the entire measured energy range.
 With these new data, different models can be restricted much more precisely,
 leading to a more accurate description of the $\gamma p\to K^0\Sigma^+$ reaction.

\section*{Acknowledgment}

The authors wish to acknowledge the excellent support of the accelerator group and
operators of MAMI. We also thank A. Sarantsev on behalf of the BnGa PWA group,
 P. Vancraeyveld and J. Ryckebusch on behalf of the RPR analysis groups,
 as well as P. Bruns for their help and constructive comments.
This work was supported by the Deutsche Forschungsgemeinschaft (SFB443,
 SFB/TR16, and SFB1044), DFG-RFBR (Grant No. 09-02-91330), the European Community-Research
Infrastructure Activity under the FP6 ``Structuring the European Research Area''
program (Hadron Physics, Contract No. RII3-CT-2004-506078), Schweizerischer
Nationalfonds, the U.K. Science and Technology Facilities Council, the U.S. Department
of Energy and National Science Foundation, and NSERC (Canada).
 A. Fix acknowledges additional support from the Russian Federation federal program
``Kadry''(Contract No. P691) and the MSE Program ``Nauka'' (Contract No. 1.604.2011).
We thank the undergraduate students of Mount Allison University
and The George Washington University for their assistance.

\end{document}